\documentstyle[prb,aps,epsfig,twocolumn]{revtex}
\begin{document}

\twocolumn[\hsize\textwidth\columnwidth\hsize\csname %
@twocolumnfalse\endcsname

\title{Density matrix renormalization group for disordered bosons in one
dimension}
\author{S.\ Rapsch\protect{\cite{perm}}, U.\ Schollw\"{o}ck and W.\ Zwerger}
\address{Sektion Physik, 
Universit\"{a}t M\"{u}nchen, Theresienstr.\ 37/III,
D-80333 M\"{u}nchen, Germany}
\date{\today}
\maketitle

\begin{abstract}
We calculate the zero-temperature phase diagram of the disordered
Bose-Hubbard
model in one dimension using the density matrix renormalization group.
For integer filling the Mott insulator is always separated from the
superfluid by a Bose glass phase. There is a reentrance of the Bose glass
both as a function of the repulsive interaction
and of
disorder. At half-filling where no Mott insulator exists, the superfluid
density has a maximum where the kinetic and repulsive energies are about
the
same. Superfluidity is suppressed both for small and very strong repulsion
but is always monotonic in disorder.
\end{abstract}
]

The interplay between disorder-induced localization and interactions
has attracted a great deal of attention in recent years. The simplest
model
including both aspects in a nutshell is a Hubbard model with random site
energies and a local repulsive interaction for either bosons or fermions
with opposite spin. Unfortunately 
there are essentially no analytical results for this model if
both
disorder and interactions are present, not even in one dimension.
For 1d bosons, however, there is a 
weak disorder, perturbative calculation by Giamarchi
and Schulz\cite{Gia}, who found that the superfluid ground state with 
quasi long range order survives disorder up to a critical point,
where the effective exponent $K$ in the decay of the one particle
density matrix is equal to $2/3$. More generally, the qualitative physics
of the Bose-Hubbard model in any dimension, and in particular the scaling 
behaviour near critical points has been elucidated by Fisher et al.\ 
\cite{Fisher}. For quantitative results, however, it is necessary
to resort to numerical simulations\cite{Scalettar,Krauth}. 
The latter were performed using path integral
(or ``world line'') Monte Carlo calculations which become increasingly
difficult in the most interesting limit of zero temperature. Now at
least
in one dimension there is an inherently zero temperature numerical
technique
for interacting quantum problems, the density matrix renormalization
group
(DMRG) method developed by White\cite{White}. It is therefore natural to
try employing this method to the disordered Bose-Hubbard model in one
dimension. This was first done by Pai et al.\cite{Pai}, who calculated
the associated phase diagram for integer filling. As expected, it
exhibits
a Mott insulating, a superfluid and also a Bose glass phase, the latter
appearing only for sufficiently strong disorder. Quite recently,
however,
their results were seriously questioned by Prokof'ev and Svistunov, who
performed rather precise quantum Monte Carlo
calculations\cite{Prokofev}.
Based on that, it was argued that the DMRG method is intrinsically
unable
to deal with disordered systems because randomness would invalidate
building
up a system in a block like fashion.

Our aim in the present work is to show that a careful DMRG calculation
can indeed be successfully applied in the presence of quenched disorder. In
particular, we provide a quantitative phase diagram for the 1d
disordered
Bose-Hubbard model at both integer and half filling. For integer filling
it is found that the superfluid and Mott insulating state are always
separated
by a Bose glass phase as suggested by Fisher et al.\cite{Fisher} The
superfluid density is nonmonotonic not only as a function of interaction
but
also of disorder. Thus for strong repulsion increasing disorder drives
a transition from a Bose glass to a superfluid.
For half filling, where no Mott insulator
exists,
the superfluid density is again a nonmonotonic function of the repulsive
interaction, however disorder now always suppresses superfluidity as
expected.
The corresponding phase diagram is in agreement with that suggested by
Giamarchi and Schulz\cite{Gia}, however we find no indication of a
qualitative difference between the glass phase at small or large values
of the repulsion (Anderson vs.\ Bose-glass).

The Bose-Hubbard model in 1d is defined by the
Hamiltonian\cite{Gia,Fisher,Scalettar}
\begin{equation}
\hat{H}= - \frac{t}{2} \sum_i (b^{\dagger}_{i+1}b_i + h.c.) +
\frac{U}{2} \sum_i n_i (n_i -1) + \sum_i \epsilon_i n_i .
\end{equation}
Here $b^{\dagger}_i$ is the boson creation operator on site $i$ of a 1d
lattice with $L$ sites and $n_i = b^{\dagger}_i b_i$ the
corresponding local occupation number with eigenvalues
$0,1,2,\ldots$. The kinetic energy is described by a hopping matrix
element $t > 0$, leading to a standard tight binding band $\epsilon (k)
= -t \cos k$
in the absence of interactions and randomness. The repulsive
interaction is described by a local, positive Hubbard $U$ which
increases the energy if more than one boson occupies a given
site. Finally the site energies $\epsilon_i$ are assumed to be
independent random variables with zero average and a box
distribution in the interval from $-\Delta$ to $\Delta$.
Throughout we work in the canonical ensemble with a given
(dimensionless) density $n = {N \over L}$ of bosons. As usual
we choose $t = 1$ as a unit of energy (note that some authors have
$t$ instead of ${t \over 2}$ in the hopping or $2 \epsilon_i$ in
the site energies which leads to a trivial factor of two
difference with our results). Apart from the density $n$, this
leaves the two dimensionless parameters $U$ and $\Delta$
characterizing the interactions and disorder which completely
specify the problem at zero temperature. In order to distinguish
the various possible phases, we calculate both the energy gap $E_g$
and the superfluid fraction $\rho_s$. The energy gap which is only
nonzero in a Mott insulating phase, can either be evaluated
directly from a numerical calculation of the energy of the ground
and first excited state. Alternatively the gap can be obtained as
the difference $E_g = \mu_p - \mu_n$ between the chemical
potential for particle ($\mu_p = E_{N+1} - E_N$) or hole
excitations\cite{Fisher} ($\mu_n = E_N - E_{N-1}$). We have
employed both methods in order to check our results. For the
superfluid fraction $\rho_s$, we use the thermodynamic definition
proposed by Fisher, Barber and Jasnow\cite{Fisher2}. It is based on defining $\rho_s$ via the
sensitivity to a change in the boundary conditions between periodic
(pbc) and antiperiodic (apbc) ones. In one dimension, at a given
density $n = {N \over L}$, the superfluid fraction $\rho_s$ is
thus given by $(t=1)$
\begin{equation}
\rho_s = {2L \over \pi^2} \cdot {L \over N} [E_0^{apbc} (L) -
E_0^{pbc} (L)]
\end{equation}
where $E_0(L)$ are the ground state energies for the specific
boundary condition. In the absence of interactions and disorder it
is straightforward to show that $\rho_s = 1$ for arbitrary
densities $n$ as it should be. It is important to note that it
is precisely a nonvanishing value of $\rho_s$ (in the limit $L
\rightarrow
\infty$) which is the relevant criterion for
superfluidity despite the fact that the one particle
density matrix $\langle b^{\dagger}_i b_0\rangle$ decays to zero
algebraically, i.e. only exhibits quasi-long range order.

For the numerical calculations it is obviously necessary to limit
the number of bosons which can occupy a given site. In order to be
able to cover also small values of $U$, where many bosons tend to
cluster at locally favorable sites, we have truncated our basis to
$m=7$ states for each site $i$ which allows up to 6 bosons
occupying the same place. We have checked carefully that our
results do not depend on $m$, which was the case at least down to $U
\approx
0.5$. 
In the DMRG calculation we studied 
system lengths up to $L = 50$ and included up to $M=190$ states. For
the truncation error, which is one minus the density matrix eigenvalues
$\lambda_\alpha$
of all $M$ states kept in the decimation, 
\begin{equation}
\rho = 1-\sum_{\alpha=1}^{M}
\lambda_\alpha ,
\end{equation}
we find values of the order of $10^{-10}$.
A very important point which turns out to be absolutely crucial
for applying the DMRG to disordered systems is to apply the finite size
(``sweeping'')
algorithm\cite{White}. After the system has been grown to its full length,
renormalization group transformations have not yet been able to take
into account the full structure of disorder while working on shorter
systems. The finite size algorithm then works on the complete system, and
improves results essentially in a variational fashion. We find good
convergence of both the gap and the superfluid fraction 
after several sweeps. The dependence on the number of
states kept was comparatively weak (also compared to the scattering of
results in various realisations of disorder) such 
that we preferred to invest computational resources
rather in sweeps.
The antiperiodic boundary condition has been implemented by
replacing the hopping energy $t$ at one of the bonds by $-t$ thus
enforcing a localized twist in the phase by $\pi$. 

\begin{figure}
\mbox{\epsfig{file=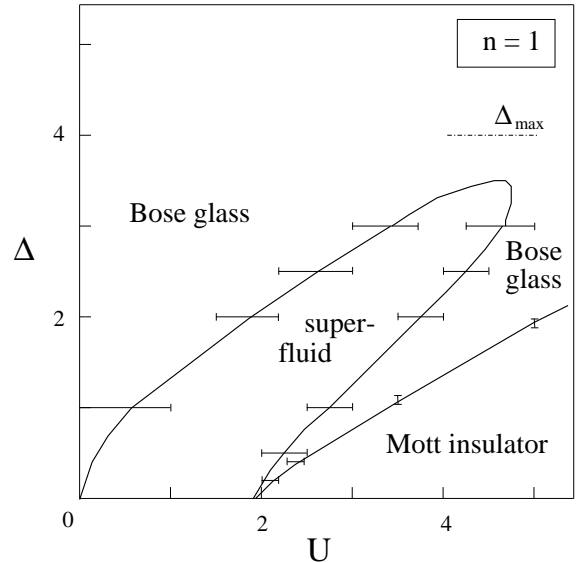,height=7.5cm}}
\vskip 5mm
\caption{Phase diagram for commensurate filling $n=1$. Error bars are
mainly due to the dependence of results on the realisations of disorder.
Above a disorder strength $\Delta_{max}=4$ 
it is always energetically advantageous to
destroy superfluidity in favor of a glass phase.}
\label{fig1}
\end{figure}

For the discussion of our results we first concentrate on a
commensurable density $n = 1$, where a Mott insulating phase is
expected\cite{Fisher} at sufficiently large $U$. In the limit of
vanishing disorder $\Delta = 0$ the system is superfluid at small
values of $U$ with a superfluid fraction $\rho_s$ which
monotonically decreases from one at $U=0$ to zero at $U=U_c$.
Since the transition to the Mott insulator is driven by phase
fluctuations at a given density, it is a Kosterlitz-Thouless like
transition\cite{Fisher} very similar to the one present in a chain
of Josephson junctions with a local charging energy\cite{Zwerger}.
Our numerical result for the critical value of $U$ is
$U_c(\Delta=0)=1.92 \pm 0.04$ which is surprisingly close to that
found in mean field theory\cite{Amico}. It also agrees with a
very recent DMRG calculation of the Bose-Hubbard model without
disorder by K\"uhner und Monien\cite{Kuehner}. They have used the
condition that the exponent $K$ characterizing the decay of the
off-diagonal density matrix
\begin{equation}
\langle b_i^+ b_0\rangle \sim |i|^{-K/2}
\label{eqcor}
\end{equation}
in the superfluid phase takes on the value $K_c = 1/2$ at the
transition\cite{Millis}. Note that $K$ scales like $\sqrt{U/t}$
at least in a Josephson junction array description which is equivalent to the 
Bose-Hubbard model at large integer densities.
\\
At finite disorder the Mott insulating phase is suppressed because
the energy gap is reduced. For vanishing hopping, i.e. $U \rightarrow
\infty$
effectively, the reduction\cite{Fisher} is just $2 \Delta$. Thus
in the limit of large $U$ the Mott-insulator disappears of $\Delta >
U/2$. This is in fact the asymptotic behaviour of the transition
line shown in Fig.\ 1. For nonzero $t$, i.e. finite $U$ the
transition appears earlier, until the Mott insulator completely
disappears at $U<U_c(\Delta = 0) = 1.92$.
Outside the Mott-insulating phase the gap vanishes, however at
finite disorder the system need not be superfluid. Indeed
calculating the superfluid fraction $\rho_s$, we find that $\rho_s$
is nonvanishing only in the superfluid regime in Fig.\ 1, which
bends down to $\Delta = 0$ both near $U=0$ and $U=U_c(\Delta=0)$.
As a consequence, at finite disorder, there is no direct transition
from a Mott insulator to a superfluid in agreement with the arguments given 
by Fisher et al.\cite{Fisher} and Freericks and Monien\cite{Freericks}.
The complete phase diagram is
shown in Fig.\ 1. It agrees well with that found by
Prokof'ev and Svistunov\cite{Prokofev} using a rather different
method and also with the qualitative picture
put forward by Herbut\cite{Herbut}. By contrast, there are strong, even qualitative
differences with the phase diagram found by Pai et al.\cite{Pai}.
Their failure to see the intervening Bose-glass between the
superfluid and the Mott-insulator is probably related to the fact
that without the sweeping algorithm the treatment of a disordered
problem by the DMRG is not reliable. 
Regarding the general structure of the phase diagram shown in
Fig.\ 1, we expect that it will not be qualitatively different for
the two dimensional case (though the corresponding
path integral Monte Carlo calculations of Krauth, Trivedi and
Ceperley\cite{Krauth} and also more recent ones\cite{Krauth1} failed to see the intermediate Bose-glass between
the superfluid and the Mott-insulator). 
Assuming that the phase
diagram of Fig.\ 1 is indeed generic for the disordered
Bose-Hubbard model at commensurate densities, one can draw two
general conclusions: 

(i) Since the superfluid fraction is a
nonmonotonic function of $U$ for a given disorder, repulsive
interactions have a delocalizing tendency at small $U$ but
enhance localization at large $U$. This is in fact a general
property, valid also at incommensurate densities as verified by
Scalettar, Batrouni and Zimanyi\cite{Scalettar} for $n = 0.625$ and
our own results at $n = 0.5$. 

(ii) More surprisingly, for fixed
repulsion in the range $U > U_c(\Delta = 0)$ but not too large,
increasing disorder drives a Bose-glass to superfluid transition.
Thus increasing disorder may in fact favour superfluidity (see
dash-dotted line in Fig.\ 1). The associated superfluid fraction is finite only
for $\Delta>\Delta_-(U)$. It first
increases with $\Delta$ but eventually decreases to zero again
at the upper boundary $\Delta_+(U)$ This
effect may be understood by observing that with increasing distance from the
Mott insulator the density of mobile particle-hole excitations increases,
thus enhancing $\rho_s$. At larger values of $\Delta$ the disorder induced 
localization takes over and $\rho_s$ goes to zero again at the upper phase 
boundary $\Delta_+ (U)$. 

\begin{figure}
\mbox{\epsfig{file=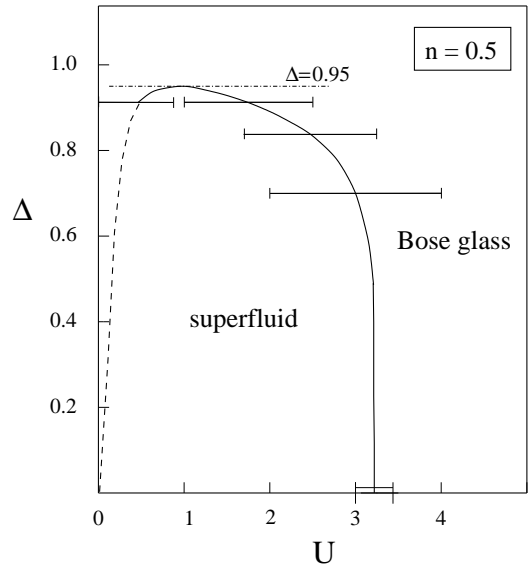,height=7.5cm}}
\vskip 5mm
\caption{Phase diagram for incommensurate filling $n=0.5$. Errors 
due to the dependence of results on the realisations of disorder are now
much stronger (with the exception of the point at $\Delta=0$).
Above a disorder strength of 0.95 we find that $\rho_S=0$ within our
numerical accuracy.}
\label{fig2}
\end{figure}

For the study of a  noncommensurate density, we have chosen $n=0.5$
where the Mott-insulating phase is completely absent\cite{Fisher}. The
resulting phase diagram is shown  in Fig.\ 2. It exhibits a superfluid
phase  in a  finite  regime $U_-(\Delta)  <  U <  U_+(\Delta)$ of  the
repulsion   provided  the   disorder   is  below   a  critical   value
$\Delta_{max} \approx  0.95$. The error  bars in the  determination of
the phase  boundary are larger than  in the case  $n=1$ because the
superfluid  fraction  exhibits   rather  strong  realization  dependent
fluctuations. This  problem becomes particularly relevant  in the limit
of small  $U$. In fact noninteracting  bosons are a  singular limit of
the disordered Bose-Hubbard model  since particles will collapse into
the single level with the lowest $\epsilon_i$, which will vary between
different  realizations. For  small but  finite $U$  the  ground state
densities are  still rather nonuniform. Now  on the basis  of that, it
has been  conjectured by  Scalettar et al.\cite{Scalettar}  that there
are two qualitatively different localized states, a suggestion 
originally due to Giamarchi and Schulz\cite{Gia}. The two phases would be
separated by a line $\Delta_c(U)$ above $\Delta_{max}$ which meets the
phase boundary to the superfluid at a multicritical point. In order
to look for signatures of this boundary at $\Delta > \Delta_{max}$,
we have calculated the expectation value of the dimensionless disorder
energy per particle 
\begin{equation}
S = \frac{1}{\Delta N} \sum_i \epsilon_i \langle n_i \rangle ,
\end{equation}
which is finite for a localized state\cite{Krauth}.
Although $S$ becomes increasingly negative as $U$ is lowered,
approaching the limit $S=-1$ at $U\ll 1$, we have found no indication
of any abrupt changes. This suggests that there is no quantitative
distinction between an ``Anderson glass'' for small $U$ and a Bose-glass for
larger repulsion. Verly likely it is only the line $U=0$ which is singular. 
This point of view is supported further by the fact that the phase diagram
found by Prokov'ev and Svistunov\cite{Prokofev} on the basis of the
Giamarchi and Schulz criterion\cite{Gia} $K_c = 2/3$ for the
renormalized exponent in the decay of the off-diagonal density
matrix (4) essentially coincides with our results. Thus for any point on the
phase boundary between the superfluid and the Bose-glass, scaling is
towards $\Delta=0, K=2/3$ even for very small $U$.
Finally we have also determined the superfluid fraction as a function of
$U$, which exhibits a maximum at $U \approx 1 - 1.5$. Unlike the case for
$n=1$ this maximum does not scale to larger $U$ if $\Delta$ is increased.
For very small $\Delta$ the critical value $U_c(\Delta=0^+) = 3.2(2)$ beyond
which $\rho_s$ vanishes in the presence of even a very small randomness has 
been determined by calculating the exponent $K$ in the {\em pure} 
system and using
the criterion $K_c = 2/3$. Quite generally, however, the
numerical calculation of $\rho_s$ becomes rather difficult for small
disorder. This is probably related to the strong divergence $\xi \sim 
\left( \frac{1}{\Delta} \right)^{1/(3-2/K)}$ of the localization length
in the limit $\Delta \rightarrow 0$ near the critical point
$K_c=2/3$, which follows from the integration
of the Giamarchi and Schulz flow equations. For vanishing disorder $\rho_s$
is finite for arbitrary values of $U$, approaching $\rho_s = 2/\pi$ as 
$U \rightarrow \infty$ where the Bose-Hubbard model at $n=\frac{1}{2}$
is equivalent to the exactly soluble quantum XY-model\cite{Mattis} in
zero magnetic field. Since $K=\infty$ in this limit, the localisation 
length in the XY-model with a random local field is expected to
diverge like $\left( \frac{1}{\Delta} \right)^{1/3}$. 

In conclusion we have demonstrated that the DMRG method can be successfully
applied to systems with quenched disorder. The phase diagram of the 1d 
Bose-Hubbard model has been determined both at integer and at half filling. 
It exhibits significant differences with earlier DMRG results\cite{Pai} but
essentially agrees with a very recent quantum Monte Carlo 
calculation\cite{Prokofev}. Our conclusions quantitatively support the general 
picture for the disordered Bose-Hubbard model 
developed by Giamarchi and Schulz\cite{Gia}
and by Fisher et al.\cite{Fisher}. The model studied here is probably the 
simplest example for the interplay between interactions and disorder 
and as such is clearly of interest in itself. Experimental realisations
e.g.\ in terms of vortices in a 1d array of Josephson 
junctions with disorder\cite{Oud} or the recent suggestion
that Bose-Hubbard physics may be relevant for cold atoms in
optical lattices\cite{BEC}, will certainly further the interest in this
model.

Acknowledgements: Useful comments by D.S. Fisher are gratefully 
acknowledged.

\end{document}